\documentclass[preprint]{emulateapj}
\usepackage{psfig,amsmath}

\shorttitle{Late time radio observations of GRB\,030329}
\shortauthors{Taylor et al.}
\begin{document}

\def\HI {H\kern0.1em{\sc i}}
\def\kipac{1}
\def\vlba{2}
\def\ar{3}
\def\cit{4}
\def\grb{GRB\,030329}
\def\simlt{\mathrel{\hbox{\rlap{\hbox{\lower4pt\hbox{$\sim$}}}\hbox{$<$}}}}
\def\simgt{\mathrel{\hbox{\rlap{\hbox{\lower4pt\hbox{$\sim$}}}\hbox{$>$}}}}

\title{~~\\ ~~\\ Late Time Observations of the Afterglow and Environment of
  GRB\,030329}

\author{G. B. Taylor \altaffilmark{\kipac,\vlba}, E. Momjian 
\altaffilmark{\ar}, Y. Pihlstr\"om\altaffilmark{\vlba,\cit}, 
T. Ghosh \altaffilmark{\ar}, \&
C. Salter \altaffilmark{\ar}  }

\altaffiltext{\kipac}{Kavli Institute of Particle Astrophysics and Cosmology,
Menlo Park, CA 94025}

\altaffiltext{\vlba}{National Radio Astronomy Observatory, Socorro,
        NM 87801}

\altaffiltext{\ar}{National Astronomy and Ionosphere Center, Arecibo Observatory, Arecibo, PR 00612}

\altaffiltext{\cit}{Owens Valley Radio Observatory, California Institute of Technology, Pasadena, CA 91125}

\begin{abstract}

We present Very Long Baseline Interferometry (VLBI) observations 217
days after the $\gamma$-ray burst of 2003 March 29.  These
observations provide further measurements of the size and position of
GRB 030329 that are used to constrain the expansion rate and proper
motion of this nearby GRB.  The expansion rate appears to be slowing
down with time, favoring expansion into a constant density
interstellar medium, rather than a circumstellar wind with an r$^{-2}$
density profile.  We also present late time Arecibo observations of
the redshifted \HI\ and OH absorption spectra towards GRB 030329.  No
absorption (or emission) is seen allowing us to place limits on the
atomic neutral hydrogen of $N_H < 8.5 \times 10^{20}$ cm$^{-2}$, and
molecular hydrogen of $N_{H_{2}} < 1.4 \times 10^{22}$ cm$^{-2}$.
Finally, we present VLA limits on the radio polarization from the
afterglow of $<$2\% at late times.

\end{abstract}

\slugcomment{As Accepted by the Astrophysical Journal}


\section{Introduction}

Our understanding of the origin of gamma-ray bursts (GRBs) has
continued to advance rapidly in the years since the first
X-ray \citep{cos97}, optical \citep{van97} and
radio \citep{fra97} afterglows were discovered.  In particular
the nearby afterglow from \grb\ has solidified the GRB-supernova connection
\citep{hjo03a,mat03}, and provided the first GRB with a well determined 
expansion rate \citep{tay04}.  This event has presented a unique 
opportunity to test afterglow models \citep{ore04, gra05}, and to explore
the environment around a GRB.

Afterglow models invoke gas-rich environments around at least some of
the GRBs. Several basic properties of this circumburst medium are
presently unknown, and need to be addressed in order to better
understand the afterglow and its evolution. For example, in the
simplest emission models, the density of the ambient medium is related
to the afterglow flux density \citep{wax97}. Especially, if GRBs exist
in both gas-poor and gas-rich environments, that could explain why
afterglow is absent in some GRBs. Direct observations of the
circumburst medium associated with GRBs would thus provide an
important test of the fireball model.

Another issue is the density profile of the medium into which the GRB
ejecta expands. Current models postulate a medium in which density is
governed by the mass-loss of the progenitor star ($\rho\propto
r^{-2}$). At present however, some observations instead point to a
uniform density in order to explain the evolution of the afterglow
lightcurve \citep{bkp+03}.  In a few cases, strong damped
Ly$\alpha$ absorption has been found indicating \HI\ column densities
as high as 5 $\times$ 10$^{21}$ cm$^{-2}$ \citep{hjo03b}.  During the
expansion of the GRB ejecta, the circumburst medium is likely to go
through different stages of ionization.
Perna \& Loeb (1998)\nocite{per98} suggest that as a consequence
(optical) absorption lines will vary in their EW with time. The rate
of this variation can constrain the size of the absorbing
region. Similarly, in the radio the fraction of molecular and atomic
gas should be governed by the effective ionization, and might
therefore vary in a similar way.

One way to investigate the circumburst medium is via atomic or
molecular absorption studies in the radio.  Another way is to
measure the deceleration of the expanding fireball.  In this
paper we present both late time Very Long Baseline Interferometry
(VLBI) observations of the 
afterglow of \grb\ and similar epoch \HI\ and OH absorption
observations taken with the Arecibo telescope.

Assuming a Lambda cosmology with $H_0 = 71$~km/s/Mpc, $\Omega_M =
0.27$ and $\Omega_\Lambda=0.73$, the angular-diameter distance of
\grb\ at $z=0.1685$ is $d_A=589\,$Mpc, and 1 milliarcsec corresponds
to 2.85 pc. 

\section{Observations and Results}\label{sec:obs}

\subsection{Late Time VLBI Observations}

\begin{deluxetable*}{lrrrrrc}
\tabletypesize{\footnotesize}
\tablecolumns{7}
\tablewidth{6in}
\tablecaption{Observational Summary\label{Observations}}
\tablehead{\colhead{Date}&\colhead{$\Delta t$}&\colhead{Frequency}&\colhead{Integ. Time}&\colhead{BW}&\colhead{Polar.}&\colhead{Instrument}\\
\colhead{}&\colhead{(days)}&\colhead{(GHz)}&\colhead{(min)}&\colhead{(MHz)}&\colhead{}&\colhead{}}
\startdata
20 Jun 2003 & 83 & 8.409 & 138 & 50 & 2 & Y27 \\
1 Nov 2003 & 217 & 8.409 & 138 & 50 & 2 & Y27 \\
1 Nov 2003 & 217 & 8.409 & 165 & 32 & 2 & VLBA+EB+Y27+WB+AR+MC+NT \\
30 Nov 2003 & 247 & 1.216 & 30 & 12 & 2 & AR \\
30 Nov 2003 & 247 & 1.428 & 50 & 12 & 2 & AR \\
30 Nov 2003 & 247 & 1.473 & 30 & 12 & 2 & AR \\
\enddata
\tablenotetext{*}{NOTE - EB = 100m Effelsberg telescope; Y27 = phased
  VLA; GBT = 105m Green Bank Telescope; WB = phased Westerbork array;
  AR = 305m Arecibo telescope; MC = Medecina 32m telescope; NT = Noto
  32m telescope.}
\end{deluxetable*}

The VLBI observations were taken at 8.4 GHz on 2003 November 1, 217
days after the burst, with a global array including the Very Long
Baseline Array (VLBA) of the NRAO\footnote{The National Radio
Astronomy Observatory is operated by Associated Universities, Inc.,
under cooperative agreement with the National Science Foundation.}.
Other telescopes used were the Effelsberg 100-m telescope\footnote{The
100-m telescope at Effelsberg is operated by the Max-Planck-Institut
f{\"u}r Radioastronomie in Bonn.}, the phased VLA, the Green Bank
Telescope (GBT), the 305 m Arecibo telescope\footnote{The Arecibo
Observatory is part of the National Astronomy and Ionosphere Center,
which is operated by Cornell University under a cooperative agreement
with the National Science Foundation.}, the Westerbork (WSRT) tied
array, and the Noto and Medecina telescopes of the Consiglio 
Nazionale delle Ricerche.  In all some
56 individual and combined antennas each of 25m or more in diameter
were employed with a combined collecting area of 0.12 km$^2$.  Most
antennas were on-source for a period of 5.5 hours.  All stations
recorded with 256 Mbps with 2 bit sampling in dual circular
polarization with the exception of Noto which had only a right-circular
polarization receiver
available.  The observations were correlated at the Joint Institute
for VLBI in Europe (JIVE).

The nearby (1.5$^\circ$) source J1051+2119 was used for
phase-referencing with a 2:1 minute cycle on source:calibrator.  The
weak calibrator J1048+2115 was observed hourly to check on the quality
of the phase referencing.  Self-calibration with a 1 hour solution
interval was used to further refine the calibration and remove some
slow-changing atmospheric phase errors.  The final image has 
an rms noise of 24 $\mu$Jy/beam.  This is substantially higher
than the expected thermal noise of 8$\mu$Jy/beam, in part
because of partial loss of signals from Arecibo
and Westerbork for reasons not completely understood. 

We fit a symmetric, two-dimensional Gaussian to the measured
visibilities on \grb\ and find a size of 0.176 $\pm$ 0.08 mas.  As in
\cite{tay04} the error of the size is estimated from signal-to-noise
ratios and from Monte-Carlo simulations of the data using identical
($u$,$v$) coverage, similar noise properties, and a Gaussian component
of known size added.  The standard deviation of the recovered sizes,
modelfitted in the same way as we treat the observations, was found to
be 0.053 mas.

We also obtain a position for \grb\ of R.A. 10$^h$44$^m$49.95955$^s$
and Dec. 21$^\circ$31'17.4377'' with an uncertainty of 0.2 mas in each
coordinate.  Solving for proper motion using all the high frequency
VLBI observations to date, we derive $\mu_{\rm r.a.}=-0.05\pm 0.41$
mas yr$^{-1}$ and $\mu_{\rm dec.}=-0.24\pm 0.41$ mas yr$^{-1}$, or an
angular displacement over the first 217 days of 0.14$\pm$0.35 mas
(Fig.~1).  These observations are consistent with those reported
by \cite{tay04}, and impose an even stronger limit on the proper motion.
This limit argues against the cannonball model for GRBs proposed by
\cite{dado04}.

\begin{figure*}
\epsscale{0.7}
\plotone{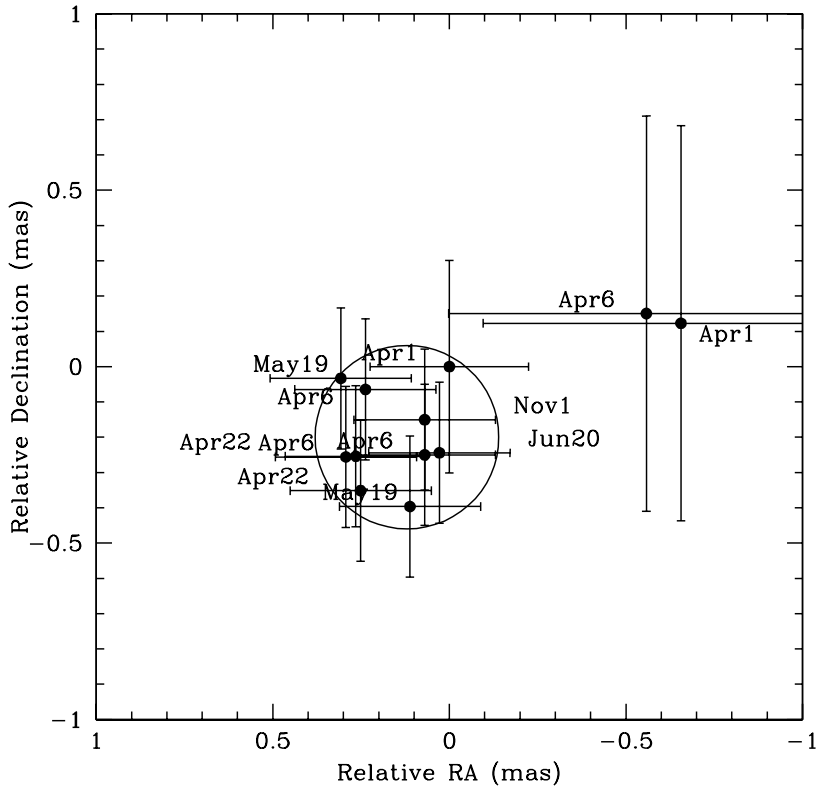}
\caption{The positions derived from the observations in six
epochs relative to the first determination on April 1st at 8.4 GHz.
Observations at multiple frequencies at a given epoch have been
plotted separately since they are independent measurements.  A circle
with a radius of 0.26 mas (2$\sigma$) is shown to encompass all
measurements except those taken at 5 GHz, which suffer from systematic
errors \citep{tay04}.  Taken together these observations provide a
constraint on the proper motion of 0.14 $\pm$ 0.35 mas over 217 days.}
\label{skypos}
\end{figure*}

\subsection{Single Dish Arecibo Observations}

Single dish observations of GRB~030329 were carried
out with the L-band Wide receiver of the 305~m Arecibo Radio
Telescope on 2003 November 30, for a total of 2~hr.
The simple position-switched observations utilized all four interim
correlator boards to observe various L-band frequencies. The bandwidth
of each board was 12.5~MHz. While the first two boards were set to
observe the frequency of the redshifted $\lambda$21~cm
H~{\footnotesize I} line, recording each linear polarization with 2048
spectral channels, the other two boards were set to observe
simultaneously the orthogonal polarizations at the frequencies of the
redshifted $\lambda$18 cm mainline OH transitions (1665 \& 1667~MHz)
and the 1720~MHz satellite OH transition with 1024 spectral channels
per polarization.

Following the editing out of data suffering from radio frequency
interference, the total on-source integration time for the redshifted
$\lambda$21~cm H~{\footnotesize I} and the 1720~MHz satellite OH
transitions was 30~min each. For the redshifted $\lambda$18 cm
mainline OH transitions, the on-source integration time was 50~min.

\section{Constraints on the Atomic and Molecular Gas}

\subsection{Limits on Absorption}

Figures~2, 3, and 4 show pairs of total-intensity spectra of the
GRB~030329 centered at the optical heliocentric velocity that
corresponds to the frequency of the redshifted $\lambda$21~cm
H~{\footnotesize I} line, the redshifted $\lambda$18~cm OH mainlines,
and the redshifted $\lambda$18~cm OH satellite line at 1720~MHz,
respectively.

The top spectra in Figs. 2, 3, and 4, are Hanning-smoothed with
spectral resolutions of 3.52~km~s$^{-1}$ (12.2~kHz), 5.99~km~s$^{-1}$
(24.4~kHz), and 5.80~km~s$^{-1}$ (24.4~kHz), respectively. The rms
noise level in each of these spectra is 1.02, 0.32, and
0.57~mJy~beam$^{-1}$, respectively.
The bottom plot in each figure is a five-channel smoothed version of
the respective top spectra. The velocity resolution in these spectra
is 8.80, 14.98, and 14.50~km~s$^{-1}$, and the rms noise level of each
spectrum is 0.482, 0.224, and 0.282~mJy~beam$^{-1}$, respectively.

No emission or absorption is seen in any of the Arecibo spectra.  We
can derive a 3$\sigma$ limit on the \HI\ opacity of $\tau < 0.53$ with
a velocity resolution of 8.8~km~s$^{-1}$.  This corresponds to a \HI\
column density limit of $N_H < 8.5 \times 10^{20}$ cm$^{-2}$ assuming
a spin temperature of 100 K, and uniform coverage.  If the spin
temperature is higher or the line is shallow and wide, then this limit
could be higher.  Given the extreme energies involved in the GRB,
enough to cause a sudden ionospheric disturbance in the Earth's
atmosphere \citep{sch03} some 740 Mpc away, it could well be that
nearly all the \HI\ along the line-of-sight in the host galaxy has
been ionized and not yet recombined.  Further research in this
area is needed.

In a similar manner we can place a 3$\sigma$ upper limit on the 1667
MHz OH mainline opacity of $\tau < 0.11$ with a velocity resolution of
15.0~km~s$^{-1}$.  This corresponds to an OH column density limit of
$N_{OH} < 1.4 \times 10^{15}$ cm$^{-2}$ assuming a uniform covering
factor, and an excitation temperature of 10 K.  We can further deduce
a limit on molecular hydrogen using the relation 
$N_{H_{2}}$ $\sim$ 10$^7$ $N_{OH}$ \citep{kan02}, of $N_{H_{2}} < 1.4 \times 
10^{22}$ cm$^{-2}$.  

The continuum flux density of the GRB~030329 at 1.4~GHz at the time of
these observations was $\sim3.5$~mJy. The various spectra reported
here show a higher continuum flux density, because of a 15~mJy
continuum source located at an angular distance of about 3~arcmin from
the GRB~030329, i.e., within the primary beam of the Arecibo radio
telescope.

\begin{figure}
\epsscale{0.7}
\plotone{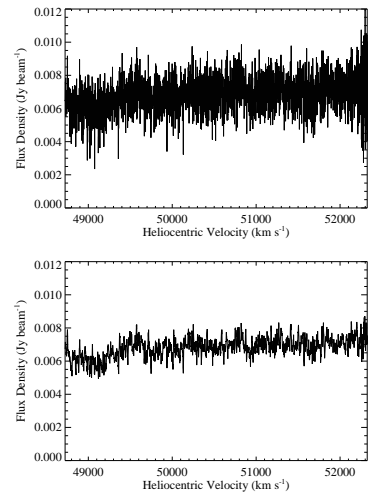}
\caption{Redshifted $\lambda$21~cm H~{\footnotesize I} spectra toward the GRB~030329. The top spectrum is Hanning-smoothed with 3.52~km~s$^{-1}$ (12.2~kHz) spectral resolution. The bottom plot is a five-channel smoothed version of the top plot, with a velocity resolution of 8.80~km~s$^{-1}$ and an rms noise level of 0.482~mJy~beam$^{-1}$.
\label{FIG1}}
\end{figure}

\begin{figure}
\epsscale{0.7}
\plotone{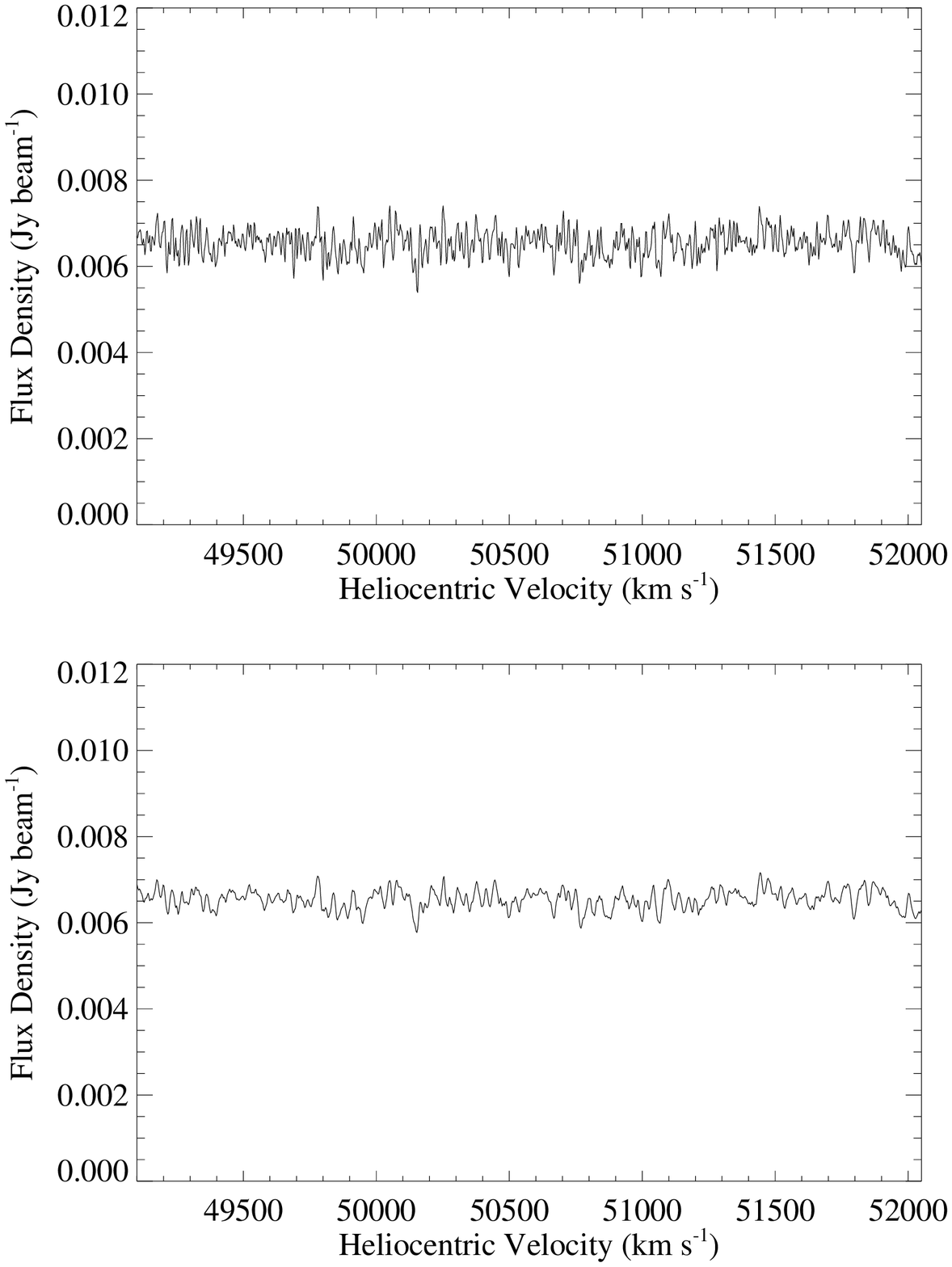}
\caption{Redshifted $\lambda$18~cm OH mainline spectra toward the GRB~030329. The top spectrum is Hanning-smoothed with 5.99~km~s$^{-1}$ (24.4~kHz) spectral resolution. The bottom plot is a five-channel smoothed version of the top plot, with a velocity resolution of 14.98~km~s$^{-1}$ and an rms noise level of 0.224~mJy~beam$^{-1}$.
\label{FIG2}}
\end{figure}

\begin{figure}
\epsscale{0.7}
\plotone{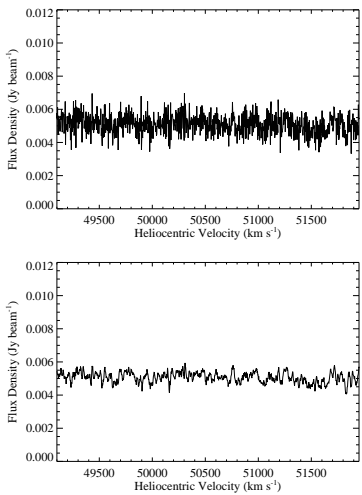}
\caption{Redshifted OH $\lambda$18~cm 1720~MHz satellite line spectra toward the GRB~030329. The top spectrum is Hanning-smoothed with 5.80~km~s$^{-1}$ (24.4~kHz) spectral resolution. The bottom plot is a five-channel smoothed version of the top plot, with a velocity resolution of 14.50~km~s$^{-1}$ and an rms noise level of 0.282~mJy~beam$^{-1}$.
\label{FIG3}}
\end{figure}

\subsection{Angular Size Measurements}\label{sec:size}

Our measured size of 0.176 $\pm$ 0.08 mas is quite close to the size
of 0.172 $\pm$ 0.043 found by \cite{tay04}, indicating a possible
slowing of the burst at late times.  This trend was already apparent
from the average expansion velocities derived from the angular size
measurements on April 22 and June 20.  The entire history of expansion
for \grb\ is shown in Fig.~5.  The first measurement at 15 days comes
from a model-dependent estimate of the quenching of the scintillation
\citep{bkp+03}. The late time curvature may indicate that between 83
and 217 days, \grb\ has transitioned into a non-relativistic
expansion.  Alternatively, it is possible (but somewhat contrived)
that the intrinsic surface brightness profile has changed in a way to
compensate for the expansion.  For all epochs we fit a two-dimensional
Gaussian model since our resolution does not permit us to 
discriminate between a Gaussian, ring, disk, or more complex profile.

\begin{figure*}
\epsscale{0.7}
\plotone{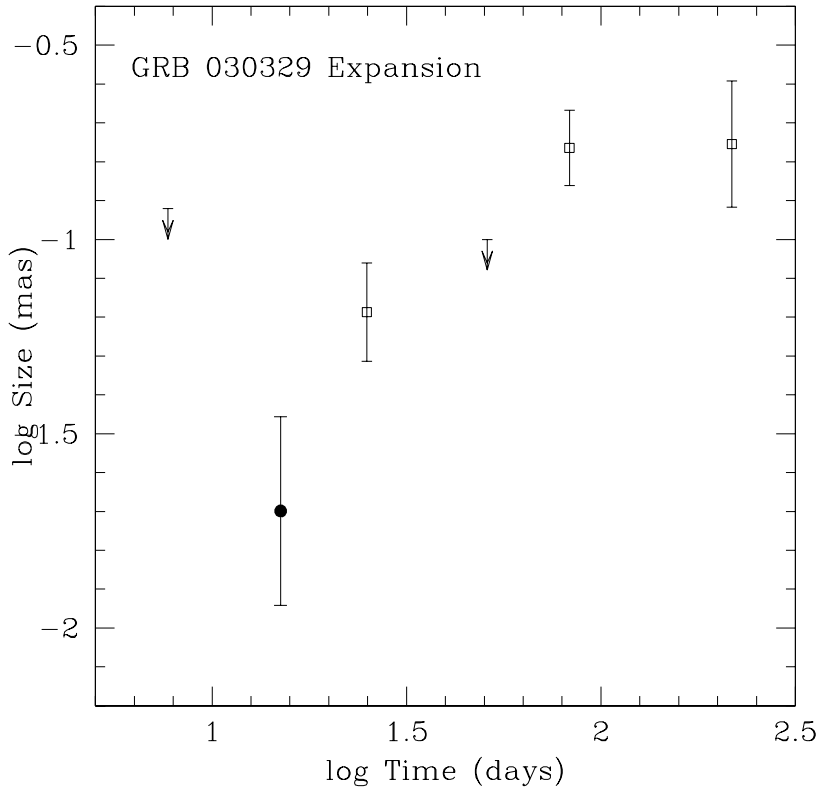}
\caption{
The apparent expansion of GRB 030329 derived from measurements and
limits on the angular size as a function of time.  The two upper
limits at 7.7 and 51 days are from \citet{tay04}, as are the
measurements on days 25 and 83 (open squares).  The measurement on day
15 (filled circle) is a model dependent estimate based on the quenching of
scintillation by \citet{bkp+03}.  Finally, the measurement on day 217 
(open square) comes from this work.}
\label{growth}
\end{figure*}

A gamma-ray burst drives a relativistic blast wave into a circumburst
medium of density $\rho$ whose radius $R$ is related to the energy of
the explosion approximately by $E\sim R^3\rho c^2\gamma^2$, where
$\gamma$ is the bulk Lorentz factor of the fireball.  More complete
treatments are given by \citep{bm76, cl00, gl03}.  The density profile
of the circumburst medium is generally taken to be either a $1/r^2$
wind, or a constant density ISM, such as one might find beyond the
termination shock of the progenitor's stellar wind \citep{che04}.  The
deceleration at late times currently favors the ISM model of
\citep{gra05} which shows a break in the expansion rate (see their
Figures 4 and 6). Unfortunately, the large uncertainty in our
measurement of the size of \grb\ at late times does not permit us to
discriminate strongly between the various models.

\cite{che04} place the termination shock of
the Wolf-Rayet wind at a distance of 0.4 pc from the progenitor.
Beyond that they predict a fairly constant density out to the red
supergiant shell at a radius of 1.7 pc.  With a current diameter of
0.5 pc for \grb\ (radius of 0.25 pc) at day 217 it is near the
termination shock, especially if the progenitor had a shorter
lifetime, or a relatively high density ISM has stalled the shock.

\subsection{Implications of low Polarization}

From 8.4 GHz VLBA observations on April 6 \cite{tay04} derived a 3$\sigma$
limit on the linear polarization of 0.16 mJy/beam, corresponding to a
limit on the fractional polarization of $<$1.0\%. In a contemporaneous
optical observation \cite{gre03} measure a polarization of
2.2+/-0.3\%.  The decrease in polarization at lower frequencies has
been explained as the result of the source being optically thick at
8.4 GHz at these early times since the maximum degree of linear
polarization of an optically thick synchrotron source is 12\% while
the maximum polarization of an optically thin synchrotron source is
about 80\% \citep{pac70}.  To look for any change in polarization with
time as the source transitions to optically thin at 8.4 GHz we have
analyzed the very sensitive, late time phased VLA observations  for
polarimetry.  We find no detection at either epoch and place 3$\sigma$
limits of $<$1.8\% and $<$4.7\% on 2003 Jun. 20 and 2003 Nov. 1
respectively.  The total intensity measured with the VLA on these
epochs is 3.11 $\pm$ 0.03 and 0.75 $\pm$ 0.02 mJy respectively.

The emission mechanism for the GRB afterglow is widely accepted to be
synchrotron radiation, which is intrinsically linearly polarized if
there is an ordered component to the magnetic field, or if the
magnetic field is generated at the internal shocks \citep{med99}.  The
recent detection of 80 $\pm$ 20\% polarization in the prompt
$\gamma$-ray emission from GRB 021206 \citep{cob03}, although
controversial \citep{rut04, wig04}, has led to an
increased interest in modeling the time and frequency dependent
behavior of the polarization from the afterglow \citep{nak03, gra03, gra05b}.

To have an observed polarization of $<$5\% in the late time afterglow
(see \S 2.2) could be explained as the result of highly disordered
magnetic fields \citep{grak03}.


Propagation effects can also reduce the intrinsic linear polarization
below detectable levels.  A Faraday screen produced by ionized gas
and magnetic fields can cause gradients in the observed polarization angle
across the source, leading to depolarization if the resolution element
of the telescope, or the size of the source if unresolved, is large compared to the gradients.   The 
rotation measures ($RM$) can be related to the line-of-sight magnetic
field, $B_{\|}$, by

$$ RM = 812\int\limits_0^L n_{\rm e} B_{\|} {\rm d}l ~{\rm
  radians~m}^{-2}~,
\eqno(1)
$$

\noindent where $B_{\|}$ is measured in mG, $n_{e}$ in cm$^{-3}$,
d$l$ in pc, and the upper limit of integration, $L$, is the distance
from the emitting source to the end of the path through the Faraday
screen along the line of sight.
We can estimate the minimum magnetic field needed to produce a
gradient across the source.  To produce a 90$^\circ$ rotation at 8.4 GHz
requires a RM of 1200 rad m$^{-2}$.  Assuming a density within
the red supergiant shell of 0.2 cm$^{-3}$ \citep{che04} and a path
length of 2 pc, then the magnetic field strength required is 
$B_{\|}=4$ mG.  This field strength is similar to the 
equipartition field strengths 
\cite{dou03} have found from modeling the radio emission from 
colliding-wind Wolf-Rayet (WR) binary systems, although the field
strengths in these systems have probably been enhanced by the collision.
Faraday screens in the GRB environment are considered in more detail
by \citet{gra05b}.

\section{Conclusions}

While no detection of atomic (\HI) or molecular (OH) material is found
towards \grb, the limits of $N_H < 8.5 \times 10^{20}$ cm$^{-2}$ are
not particularly constraining.  Observations of more shielded
molecules like NH$_3$ towards future bursts at early times would be of
interest, and given the expected strength of a nearby afterglow of
$\sim$50 mJy or more, could yield detections, or place interesting
limits on the amount of material in the GRB environment.

Although GRB 030329 has faded considerably, it may still be detectable
with VLBI techniques.  Even crude estimates of the size could
differentiate between the predictions of wind and constant-density
environments at these late times.  Radio re-brightening of
\grb\ has been predicted by \cite{gl03}, and \cite{li04} estimate a
level of 0.6 mJy 1.7 years after the burst.  This re-brightening might
occur as the counterjet becomes non-relativistic and therefore
radiatively isotropic.  This brightening will be accompanied by a
temporary rapid growth in the size of the source as two,
well-separated jets become visible, and by a change in the light
centroid \citep{gl03}.  If such a re-brightening occurs then a precise
size estimate at late times becomes readily achievable with existing
facilities.

Rather surprisingly, we find no detectable linear polarization from
\grb\ at cm wavelengths to limits as low as 1\%.  This could indicate
less order than expected in the magnetic fields of the external shock
that drives the afterglow, or a Faraday screen that depolarizes the
radio emission.  All of these limits are from 8 days or more after the
burst.  Given the RHESSI result \citep{cob03} of large polarization
from the $\gamma$-rays, and predictions of some fireball models, it
would be well worth searching for polarization from the prompt optical
and radio emission ({\it e.g.}, Granot \& Taylor 2005\nocite{gra05b}).

In the near future the {\it Swift} satellite\footnote{see 
http://swift.gsfc.nasa.gov} should dramatically increase the
number of GRBs with measured redshifts, revealing some that
are nearby.  
In future VLBI studies of GRB afterglows at redshifts less
than 0.1 it should be possible to image the structure of
the afterglow.
Fireball models of heating by a single relativistic shock front
predict that at late times the fireball should look like a ring
\citep{gps99b}.  

\acknowledgments

GBT thanks the Kavli Institute for Particle Astrophysics and
Cosmology for hospitality and support.  We thank
Jonathon Granot and Avi Loeb for useful discussions.  This
research has made use of NASA's Astrophysics Data System.


\begin{thebibliography}{}

\bibitem[{Berger} {\it et al.}\ (2003)]{bkp+03}
{Berger}, E. {\it et al.}\  2003, Nature, 426, 154


\bibitem[Blandford \& McKee(1976)]{bm76}
Blandford, R.~D. and McKee, C.~F. 1976, Phys. of Fluids, 19, 1130

\bibitem[{Chevalier} \& {Li}(2000)]{cl00}
{Chevalier}, R.~A. and {Li}, Z. 2000, ApJ, 536, 195

\bibitem[{Chevalier, Li, \& Fransson}(2004)]{che04}
{Chevalier}, R.~A., {Li}, Z., \& Fransson, C. 2004, ApJ, 606, 369

\bibitem[Coburn \& Boggs(2003)]{cob03}
Coburn, W., \& Boggs, S.~E. 2003, Nature, 423, 415

\bibitem[Costa et al.(1997)]{cos97} Costa, E., et al.\ 1997, 
Nature, 387, 783 

\bibitem[Dado, Dar \& De~Rujula(2004)]{dado04}
Dado, S., Dar, A., and De~Rujula, A. 2004, astro-ph/0402374.

\bibitem[Dougherty et al.(2003)]{dou03} Dougherty, S.~M., 
Pittard, J.~M., Kasian, L., Coker, R.~F., Williams, P.~M., \& Lloyd, H.~M.\ 
2003, A\&A, 409, 217 

\bibitem[{Frail} {\it et al.}\ (1997)]{fra97}
{Frail}, D.~A., {Kulkarni}, S.~R., {Nicastro}, S.~R., {Feroci}, M., and
  {Taylor}, G.~B. 1997, Nature, 389, 261


\bibitem[{Granot}, {Piran} \& {Sari}(1999)]{gps99b}
{Granot}, J., {Piran}, T., and {Sari}, R. 1999, ApJ, 513, 679

\bibitem[Granot(2003)]{gra03} Granot, J.\ 2003, \apjl, 596, L17 

\bibitem[Granot \& K{\" o}nigl(2003)]{grak03} 
Granot, J.~\& K{\" o}nigl, A.\ 2003, ApJ, 594, L83 

\bibitem[{Granot} \& {Loeb}(2003)]{gl03}
{Granot}, J. and {Loeb}, A. 2003, ApJ, 593, L81

\bibitem[Granot, Ramirez-Ruiz \& Loeb(2005)]{gra05} 
Granot, J., Ramirez-Ruiz, E., \& Loeb, A.\ 2005, ApJ, in press, 
astro-ph/0407182

\bibitem[Granot \& Taylor(2005)]{gra05b} 
Granot, J., \& Taylor, G.B.\ 2005, ApJ, submitted, 
astro-ph/04120309

\bibitem[{Greiner}{\it et al.}(2003)]{gre03} 
{Greiner}, J., et al.\ 2003, Nature, 426, 157

\bibitem[Hjorth et al.(2003a)]{hjo03a} Hjorth, J., et al.\ 
2003a, \nat, 423, 847 

\bibitem[Hjorth et al.(2003b)]{hjo03b} Hjorth, J., et al.\ 
2003b, \apj, 597, 699 

\bibitem[Kanekar \& Chengalur(2002)]{kan02}
Kanekar, N., \& Chengalur, J.~N. 2002, A\&A, 381, L73

\bibitem[Li \& Song(2004)]{li04} Li, Z.~\& Song, L.~M.\ 
2004, \apjl, 614, L17 

\bibitem[Matheson et al.(2003)]{mat03} Matheson, T., et al.\ 
2003, \apj, 599, 394 

\bibitem[Medvedev \& Loeb(1999)]{med99} Medvedev, M.~V.~\& 
Loeb, A.\ 1999, ApJ, 526, 697 


\bibitem[Nakar, Piran \& Waxman(2003)]{nak03}
Nakar, E., Piran, T., \& Waxman, E. 2003, JCAP, 10, 5

\bibitem[Oren, Nakar, \& Piran(2004)]{ore04} Oren, Y., Nakar, 
E., \& Piran, T.\ 2004, MNRAS, 353, L35 

\bibitem[Pacholczyk(1970)]{pac70}
Pacholczyk, A. G. 1970, Radio Astrophysics: Nonthermal Processes in Galactic and Extragalactic Sources (New York: Freeman)

\bibitem[van Paradijs et al.(1997)]{van97} van Paradijs, J., 
et al.\ 1997, Nature, 386, 686 

\bibitem[Perna \& Loeb(1998)]{per98} Perna, R., \& Loeb, A.\ 1998, 
ApJ, 501, 467

\bibitem[Rutledge \& Fox(2004)]{rut04}
Rutledge, R.~E., \& Fox, D.~B. 2004, MNRAS, 350, 1288

\bibitem[Schnoor {\it et al.}(2003)]{sch03}
Schnoor, P.~W., Welch, D.~L., Fishman, G.~J., \& Price, A.\ 2003,
GCN Circ. No. 2176

\bibitem[Taylor {\it et al.}(2004)]{tay04}
Taylor, G.~B., Frail, D.~A., Berger, E., and Kulkarni, S.~R. 2004, ApJ,
  609, L1

\bibitem[Waxman(1997)]{wax97}
Waxman, E. 1997, ApJ, 491, L19

\bibitem[Wigger et al.(2004)]{wig04}
Wigger, C., et al. 2004, ApJ in press (astro-ph/0405525)


\end{thebibliography}
\end{document}